\documentclass[11pt]{article}

\usepackage{latexsym,amsmath,amssymb,amscd,eucal}
\usepackage{xypic}
\usepackage{enumerate}

\def\harr#1#2{\smash{\mathop{\hbox to .3in{\rightarrowfill}}
 \limits^{\scriptstyle#1}_{\scriptstyle#2}}}

\def\appendix#1{\addtocounter{section}{1}\setcounter{equation}{0}
\renewcommand{\thesection}{\Alph{section}}
\section*{Appendix \thesection\protect\indent \parbox[t]{11.715cm} {#1}}
\addcontentsline{toc}{section}{Appendix \thesection\ \ \ #1} }
\newcommand{\eq}{\begin{equation}}
\newcommand{\eqend}{\end{equation}}

\newbox\ncintdbox \newbox\ncinttbox
\setbox0=\hbox{$-$} \setbox2=\hbox{$\displaystyle\int$}
\setbox\ncintdbox=\hbox{\rlap{\hbox to
\wd2{\hskip-.125em\box2\relax \hfil}}\box0\kern.1em}
\setbox0=\hbox{$\vcenter{\hrule width 4pt}$}
\setbox2=\hbox{$\textstyle\int$}
\setbox\ncinttbox=\hbox{\rlap{\hbox to
\wd2{\hskip-.175em\box2\relax \hfil}}\box0\kern.1em}

\def\be{\begin{equation}}
\def\ee{\end{equation}}
\def\bea{\begin{eqnarray}}
\def\eea{\end{eqnarray}}
\def\bd{\begin{displaymath}}
\def\ed{\end{displaymath}}

\DeclareFontFamily{U}{rsf}{}
\DeclareFontShape{U}{rsf}{m}{n}{
  <5> <6> rsfs5 <7> <8> <9> rsfs7 <10-> rsfs10}{}
\DeclareMathAlphabet\Scr{U}{rsf}{m}{n}

\makeatletter
\newdimen\normalarrayskip              
\newdimen\minarrayskip                 
\normalarrayskip\baselineskip
\minarrayskip\jot
\newif\ifold             \oldtrue            
\def\arraymode{\ifold\relax\else\displaystyle\fi} 
\def\@arrayskip{\ifold\baselineskip\z@\lineskip\z@
     \else
     \baselineskip\minarrayskip\lineskip2\minarrayskip\fi}
\def\@arrayclassz{\ifcase \@lastchclass \@acolampacol \or
\@ampacol \or \or \or \@addamp \or
   \@acolampacol \or \@firstampfalse \@acol \fi
\edef\@preamble{\@preamble
  \ifcase \@chnum
     \hfil$\relax\arraymode\@sharp$\hfil
     \or $\relax\arraymode\@sharp$\hfil
     \or \hfil$\relax\arraymode\@sharp$\fi}}
\def\@array[#1]#2{\setbox\@arstrutbox=\hbox{\vrule
     height\arraystretch \ht\strutbox
     depth\arraystretch \dp\strutbox
     width\z@}\@mkpream{#2}\edef\@preamble{\halign \noexpand\@halignto
\bgroup \tabskip\z@ \@arstrut \@preamble \tabskip\z@ \cr}%
\let\@startpbox\@@startpbox \let\@endpbox\@@endpbox
  \if #1t\vtop \else \if#1b\vbox \else \vcenter \fi\fi
  \bgroup \let\par\relax
  \let\@sharp##\let\protect\relax
  \@arrayskip\@preamble}
\makeatother

\newcommand{\beq}{\begin{eqnarray}}
\newcommand{\eeq}{\end{eqnarray}}

\def\appendix#1{\addtocounter{section}{1}\setcounter{equation}{0}
\renewcommand{\thesection}{\Alph{section}}
\section*{Appendix \thesection. #1}
\addcontentsline{toc}{section}{Appendix \thesection\ \ \ #1} }

\numberwithin{equation}{section}

\begin{document}


\vspace{.1in}

\begin{center}

{\Large\bf EXPOSITORY REMARKS ON THREE-DIMENSIONAL GRAVITY AND HYPERBOLIC INVARIANTS}

\end{center}
\vspace{0.1in}
\begin{center}
{\large
A. A. Bytsenko $^{(a)}$\footnote{ abyts@uel.br}\,
and M. E. X. Guimar\~aes $^{(b)}$
\footnote{emilia@if.uff.br}}
\vspace{7mm}
\\
$^{(a)}$ {\it Departamento de F\'{\i}sica, Universidade Estadual de Londrina\\
Caixa Postal 6001, Londrina-Paran\'a, Brazil}
\vspace{5mm}\\
$^{(b)}$ {\it Instituto de F\'{\i}sica,
Universidade Federal Fluminense,\\
Av. Gal. Milton Tavares de Souza s/n, Niter\'oi-RJ, Brazil}
\vspace{5mm}\\

\end{center}
\vspace{0.1in}
\begin{center}
{\bf Abstract}
\end{center}
We consider complex invariants associated with compact real
three-dimensional hyperbolic spaces. The contribution of the
Chern-Simons invariants of irreducible $U(n)-$flat connections on
hyperbolic fibered manifolds to the low order expansion of the
quantum gravitational path integral is analyzed.

\vfill




\newpage



\section{Introduction}

In the Euclidean path integral approach to quantum gravity the
partition function can be presented as a weighted sum over
four-space \cite{Hawking}. Computation of the partition function
is not easy to implement because of non-renormalizability of
general relativity, unbounded from below the action, and problem
of classification of relevant geometries. Three-dimensional
gravity provides a simple model which is renormalizable at the
quantum level, and its relationship to Chern-Simons theories
makes a systematic perturbation expansion possible.

In this note
we consider (up to the first order of loop expansion) the
contribution to the gravitational path integral associated with three-dimensional real hyperbolic
spaces. Let $X = G/{K}$ be an irreducible rank one
symmetric space of non-compact type. Thus $G$ will be a connected
non--compact simple split rank one Lie group with finite center,
and ${K}\subset G$ will be a maximal compact subgroup.
The object of interest is the groups $G=SO_1(3,1)$ and ${K}=SO(3)$. The corresponding
symmetric space of non-compact type is the real hyperbolic space
$X = {\mathbb H}^3 = SO_1(3, 1)/SO(3)$ of sectional curvature
$-1$.
Gravitational path integral in the Euclidean sector has the form
\begin{equation}
Z(X) = \int [d{g}] \exp\left\{ {(16\pi G)^{-1}}
\int_{X} d^3\!x
\sqrt{{g}}(R[{g}]-2\Lambda)\right\}
\mbox{,}
\label{action}
\end{equation}
where the integration is over all Riemannian metrics $g$
on a three-dimensional space $X$, $R[g]$ is
the scalar curvature and $\Lambda$ is the cosmological constant.
The extrema of the action (\ref{action}) are Einstein spaces
for which
$
R_{\mu\nu}[g] = 2\Lambda g_{\mu\nu}.
$
In three dimensions the Ricci tensor completely determines the full
curvature, it means that the Einstein space always has constant
curvature. Thus, an extremum $(X,g)$ of the action must be an elliptic
($\Lambda>0$), hyperbolic ($\Lambda<0$), or flat ($\Lambda=0$) space.
In the neighborhood of an extremum of (\ref{action}) the semiclassical
approximation gives (see, for example, \cite{Carlip2})
\begin{equation}
Z_{sc}(X) =  \left| \pi_0(\mbox{\it Diff}\,X )\right|^{-1}
\sum_{\mbox{\scriptsize extrema}}D(X)\exp\left\{
\mbox{sgn}(\Lambda) \frac{{\rm Vol}(X)}{4\pi
G|\Lambda|^{1/2}}\right\} \mbox{,} \label{part}
\end{equation}
where $|\pi_0(\mbox{\it Diff}\,X)|$ is the order of the mapping
class group of $X$, and ${\rm Vol}(X)$ is the volume of $X$
with the constant curvature metric.
We comment here a possible role and contribution of the sum over
topologies of invariants associated with hyperbolic geometry
(the analytic torsion with its relation to the Chern--Simons functional)
which might be significant in three--dimensional quantum gravity.
Note also the recent
data obtained by the Wilkinson microwave anisotropy probe (WMAP)
\cite{wmap} satellite confirmed, and set new standards of
accuracy to the previous COBE's measurement of a low quadrupole
moment in the angular power spectrum of the CMB, which is in
accordance with the assumption that the topology of the universe
might be non--trivial, with particular emphasis on the case of a
hyperbolic universe.

\section{Complex topological invariants}

The prefactor $D(X)$ in (\ref{part}) is a combination of
determinants coming from small fluctuations around $g$ and from
gauge fixing; it can be computed by taking in the connection
between three-dimensional gravity and Chern-Simons theory
\cite{Witten}. Let us consider the classical Chern-Simons
functional $CS(X_{\Gamma})$ which is a function on a space of
connections $A$ on bundles over a compact oriented three-manifold
$X_{\Gamma}$ where $\Gamma$ is a co-compact subgroup of $G$. The
value of the function $CS(X_{\Gamma})$ on the space of connections
${\mathfrak A}_{X_{\Gamma}},\, A\in{\mathfrak A}_{X_{\Gamma}}$, at
a critical point can be regarded as a topological invariant of a
pair $(X_{\Gamma},\chi)$, where $\chi$ is an orthogonal
representation of $\Gamma$. If a smooth four-manifold $M$ has a
boundary $\partial M\equiv X_{\Gamma}$, the index of a Dirac
operator ${\mathfrak D}_{{A}}$ acting on $M$ is given by
Atiyah-Patodi-Singer theorem \cite{Atiyah1,Atiyah2,Atiyah3}
\begin{equation}
{\rm Index}({\mathfrak D}_{{A}})=\int_{M}{\rm ch}(M)
{\widehat A}(M)- \frac{1}{2}[\eta(0,{\mathfrak D}_{\chi})+
h(0,{\mathfrak D}_{\chi})],
\end{equation}
where the Chern character ${\rm ch}(M)$ and the
${\widehat A}(M)-$genus are the usual polynomial related to Riemannian
curvature of $M$, $h(0,{\mathfrak D}_{\chi})$ is the dimension of
the space of harmonic spinors on
$X_{\Gamma}$ ($h(0,{\mathfrak D}_{\chi})
={\rm dim}{\rm Ker}\,{\mathfrak D}_{\chi}$ =
multiplicity of the 0-eigenvalue of ${\mathfrak  D}_{\chi}$ acting on
$X_{\Gamma}$); ${\mathfrak D}_{\chi}$ is a Dirac operator on
$X_{\Gamma}$ acting on spinors with coefficients in $\chi$.
The Chern-Simons invariant of
$X_{\Gamma}=\Gamma\backslash{\mathbb H}^3$ can be derived
from the index of the Dirac operator. The following formula for the
$U(n)-$Chern-Simons invariant of an irreducible flat connection
on the real hyperbolic three-manifolds holds \cite{Bonora}:
\begin{equation}
CS_{U(n)}(X_{\Gamma}) = \frac{1}{2\pi {\sqrt{-1}}}{\rm log}
\left[\frac{Z(0,{\mathfrak D})^{{\rm dim}\chi}} {Z(0,{\mathfrak
D_{\chi}})} \right]\,\,\,\,\,\,\,\,\,\, {\rm modulo}\, ({\mathbb
Z}/2) \mbox{,} \label{CSF}
\end{equation}
where $Z(s,{\mathfrak D}_{\chi})$ is the Selberg-type (Shintani)
spectral function.
It gives the explicit form of the prefactor in
semiclassical approximation to the Chern-Simons theory. The first
quantum correction to the partition function associated with the
Chern--Simons gauge theory has the form ${W}(k)=\int
[DA]e^{{\sqrt{-1}}kCS(X_{\Gamma})},\, k\in {\mathbb Z}.$ The
function $W(k)$ can be evaluated in terms of $L^2-$analytic
torsion $[T_{an}(X_{\Gamma})]^2$, the spectral zeta function
$\zeta(s,|{\mathfrak D}|)$ and the spectral function
$Z(s,{\mathfrak D}_{\chi})$. The final result is \cite{Bytsenko4}
\begin{equation}
{W}(k)=\left(\frac{\pi}{k}\right)^{\zeta(0,|{\mathfrak D}|)/2}
{Z}(0,{\mathfrak D})^{-1/4}
\left[T_{an} (X_\Gamma)\right]^{1/2}
\left[{\rm Vol}(\Gamma\backslash G)\right]^{-{\rm dim}H^0(\nabla)/2}
\mbox{.}
\end{equation}
The analytic torsion for manifold $X_\Gamma$ has been calculated
(in the presence of non-vanishing Betti numbers $b_i\equiv
b_i(X_{\Gamma}$)) in \cite{Bytsenko4}, and it is given by
\begin{equation}
[T_{an}(X_{\Gamma})]^2 = \frac{(b_1-b_0)![Z(2, b_0)]^2}
{[b_0!]^2Z(1, b_1-b_0)} \exp\left(-\frac{1}{3\pi}{\rm
Vol}(\Gamma\backslash G)\right) \mbox{.} \label{torsion}
\end{equation}

{\bf Lower K-groups and holonomy invariants.}
Let as before $X_{\Gamma}$ be a real compact oriented three dimensional
hyperbolic manifold. Its fundamental group $\Gamma$ comes with map
to $PSL(2, {\mathbb C})\equiv SL(2, {\mathbb C})/\{\pm Id\}$;
therefore in general one gets a class in group homology
$H_3(GL({\mathbb C}))$. For algebraic K-groups the following result
hold \cite{Suslin1,Suslin3}: for a field ${\mathcal F}$,
\begin{equation}
K_j({\mathcal F})\cong
H_j(GL(j, {\mathcal F}))/H_j(GL(j-1, {\mathcal F})
\mbox{.}
\end{equation}
The group $K_3({\mathcal F})$ is built out of $K_3({\mathcal F})$
and the Bloch group ${\mathfrak B}({\mathcal F})$.
There is an exact sequence due to Bloch and Wigner:
\begin{equation}
0\longrightarrow {\mathbb Q}/{\mathbb Z}\longrightarrow H_3(PSL(2,
{\mathbb C}); {\mathbb Z})\longrightarrow {\mathfrak B}({\mathbb
C})\longrightarrow 0\,.
\end{equation}
The Bloch group ${\mathfrak B}({\mathbb C})$ is known to be
uniquely divisible, so it has canonically the structure of a
${\mathbb Q}-$vector space \cite{Suslin87}. The ${\mathbb
Q}/{\mathbb Z}$ in the Bloch-Wigner exact sequence is precisely
the torsion of $H_3(PSL(2, {\mathbb C}); {\mathbb Z})$.
Since we are looking for $H_3(GL(2, \bullet))$, the homology invariant
of a hyperbolic three-manifold should live in the Bloch group
${\mathfrak B}(\bullet)$. The following result confirmed this
statement. A real oriented finite-volume hyperbolic three-manifold
$X = X_{\Gamma}$ has an invariant \cite{Neumann1}:
$\beta (X) \in {\mathfrak B}({\mathbb C})$.
Actually $\beta (X)\in {\mathfrak B}({\mathcal F})$ for an associated
number field ${\mathcal F}(X)$. In fact, under the (normalized) Bloch
regulator
${\mathfrak B}({\mathbb C})\rightarrow {\mathbb C}/{\mathbb Q}$,
invariant $\beta (X)$ goes to
$\{(2/\pi){\rm vol} (X) +4\pi\sqrt{-1}CS(X)\}$.
Let assume that ${\mathcal F}(X)$ can be embedded in ${\mathbb C}$ as
an imaginary quadratic extension of a totally real number field,
then $CS(X)$ is rational. Conjecturally, $CS(X)$ is irrational if
${\mathcal F}(X)\cap \overline{{\mathcal F}(X)}\subset {\mathbb R}$
\cite{Rosenberg1}.

{\bf Thurston's suggestion.}
Recall that in two dimensional quantum gravity, as defined by
string theories, the sum over all topologies can be performed in
the functional integral formalism: $\int [dg] =\sum_{g=0}^
{\infty}\int_{(\rm genus)}[dg]$. A necessary first step to
implement this formalism to the three dimensional case is the
classification of all possible three geometries  $(X,g)$ by
Kleinian groups \cite{Bytsenko0}. If Thurston's conjecture is true \cite{Thurston},
then every compact closed three dimensional manifold can be
represented as follows:
$\bigcup_{\ell=1}^{\infty}(\Gamma_{n_{\ell}} \backslash
G_{n_{\ell}})$, where $n_{\ell} \in (1,...,8)$ corresponds to one
of the eight geometries, and $\Gamma$ is the (discrete) isometry
group of that geometry.  It has to be noted that gluing the above
geometries, characterizing different coupling constants by a
complicated set of moduli, is a very difficult task. Perhaps this
can be done, however, with a bit of luck.

Important geometric invariants which can be defined are the volume
and the Chern--Simons invariants. Thurston suggests to combine
these two invariants into a single complex invariant whose
absolute value is $\exp\left((2/\pi){\rm Vol}(X)\right)$ and whose
argument is the Chern-Simons invariant of $X$, $CS(X)$
\cite{Thurston}. Taking into account the Thurston's classification
of all possible three geometries and $\beta(X)$ (see previous
section) this invariant can also be presented in the form
\begin{equation}
{\mathfrak W}=
\exp\left\{\bigcup_{\ell=1}^{\infty}
\left(\frac{2}{\pi}
{\rm Vol}(\Gamma_{n\ell}\backslash G_{n\ell})+
4\pi \sqrt{-1} CS(\Gamma_{n\ell}\backslash G_{n\ell})\right)
\right\}
\mbox{.}
\label{Thurston}
\end{equation}
If we make the intuitive requirement that only irreducible
manifolds have to be taken into account (for instance,
supersymmetry surviving arguments in favor of this requirement
\cite{Bytsenko22}), then the manifolds modelled on $ {\mathbb
S}^2\times {\mathbb R}, \,{\mathbb H}^2\times {\mathbb R} $ have
to be excluded from Thurston's list. There is only a finite number
of manifolds of the form $\Gamma \backslash {\mathbb R}^N$,
$\Gamma \backslash{\mathbb S}^N$ for any $N$ \cite{Wolf}. Perhaps
the more important contribution to the vacuum persistence
amplitude should be given by the hyperbolic geometry, the other
geometries appearing only for a small number of exceptions
\cite{Besse}.

{\bf The Haken's class.}
Going back to the calculation of sums over topologies, we note the
particular problem of classification of manifolds of a given
class. A solution of this task lies in producing of two
algorithms. First of them, the numeration algorithm, must
enumerate, possibly with repetitions, the manifolds of a given
class. The second one, the recognition algorithm, applied to any
two given manifolds of the class in question, must tell us whether
or not they are homeomorphic. The joint action of these algorithms
leads to the appearance of an infinite list which contains all
manifolds of the given class without duplication, and therefore
the manifolds can be described algorithmically. Such a solution of
the classification problem can be regarded as satisfactory only up
to a first approximation (in a weak sense), usually mathematicians
require more. In general, hyperbolic manifolds have not been
completely classified and therefore a systematic computation is
not yet possible. However it is not the case for sufficiently
large manifolds \cite{Haken}, which give an essential contribution
to the torsion (\ref{torsion}). There is a class of three
dimensional sufficiently large hyperbolic manifolds which admits
arbitrary large value of $b_1(X)= {\rm rank}_{\mathbb
Z}H_1(X;{\mathbb Z})$. Sufficiently large manifold contains a
surface $\Sigma$ whereas $\pi_1(\Sigma)$ is finite and
$\pi_1(\Sigma) \subset \pi_1(X)$. It is known that any
three-manifold can be triangulated, and hence can be partitioned
into handles. Since the existence and uniqueness of a
decomposition of an orientable manifold as the sum of simple
orientable parts have been established (see \cite{Milnor,Hempel}),
the question of homeomorphy can be considered only for irreducible
manifolds. The method proposed by Haken \cite{Haken} permits to
describe all normal surfaces of a three-manifold $X$ which has
been partitioned on handles previously. The Haken's theory of normal
surfaces was further verified for the procedure of geometric
summation of surfaces. As a result, the classification theorem
says \cite{Matveev}:

$({\bf i})$ There exists an algorithm for
enumerating  all of the Haken manifolds;

$({\bf ii})$ There exists
an algorithm for recognizing homeomorphy of the Haken manifolds.


\section*{Acknowledgements}

A. A. Bytsenko and M. E. X. Guimar\~aes would like to thank the
Conselho Nacional de Desenvolvimento Cient\'{\i}fico e
Tecnol\'ogico (CNPq, Brazil) for partial support.

\end{document}